\documentclass[12pt]{article}
\input psfig

\renewcommand{\section}[1]{\vspace{6pt} \noindent\mbox{#1} \newline \noindent}
\renewcommand{\subsection}[1]{\vspace{6pt} \noindent\mbox{\underline{#1}} 
\newline \noindent}
\renewcommand{\subsubsection}[1]{\vspace{6pt} \noindent\mbox{\underline{#1}}
\noindent}

\newfont{\sansb}{cmssbx10}
\newfont{\sans}{cmss10}

\newcommand{\vol}[2]{$\;\,$\bf #1\rm , #2.} 
\def\aaps{{\it Astron. Astr. Suppl.}}                 
\def\ssr{{\it Space Sci. Rev.}}                       
\def\apj{{\it Astrophys. J.}}                         
\def\apjl{{\it Astrophys. J. (Lett.)}}                
\def\mnras{{\it MNRAS}}                               
{\catcode`\@=11                                                  
\gdef\SchlangeUnter#1#2{\lower2pt\vbox{\baselineskip 0pt\lineskip0pt    
\ialign{$\m@th#1\hfil##\hfil$\crcr#2\crcr\sim\crcr}}}}

\def\gray{$\gamma$-ray}
\def\grays{$\gamma$-rays}
\def\etal{{\it et al.}}

\begin{document}
{\small OG 3.4.4 \vspace{-24pt}\\}     

{\center \LARGE GAMMA-RAY ABSORPTION AT HIGH REDSHIFTS AND THE GAMMA-RAY 
BACKGROUND \vspace{14pt}\\}

F. W. Stecker$^{1}$ and M. H. Salamon$^2$ 
   \vspace{2pt}\\

{\it $^1$LHEA, NASA/Goddard Space Flight Center, Greenbelt, MD 20770, USA\\
$^2$Physics Dept., University of Utah, Salt Lake City, UT 84112, USA\vspace{-12pt}\\}

{\center ABSTRACT\vspace{3pt}\\}
We present results of a calculation of absorption of 10-500 GeV $\gamma$-rays
at high redshifts (Salamon and Stecker, 1997).
This calculation requires the determination of the high 
redshift evolution of the full spectral energy distribution of the 
intergalactic photon field. For this, we have primarily followed the recent
analysis of Fall, Charlot \& Pei. We give our results for the $\gamma$-ray
opacity as a function of redshift out to a redshift of 3. We then give 
predicted $\gamma$-ray spectra for selected blazars and also extend our results
on the background from unresolved blazars to an energy of 500 GeV. Absorption 
effects are predicted to significantly steepen the background spectrum above 
20 GeV. Our absorption calculations can be used to place limits on the 
redshifts of $\gamma$-ray bursts. Our background calculations can be used to 
determine the observability of multi-GeV lines from dark matter (neutralino) 
particles.
\setlength{\parindent}{1cm}

\section{INTRODUCTION}   
Absorption of $\gamma$-rays from blazars and extragalactic $\gamma$-ray bursts
is strongly dependent on the redshift of the source (Stecker, De Jager \&
Salamon 1992). Stecker \& De Jager (1997) have calculated the absorption of
$\gamma$-rays at above 0.3 TeV at redshifts up to 0.54. The study of 
extragalactic absorption below 0.3 TeV at higher redshifts is a more complex
and physically interesting subject. In order to calculate such absorption 
properly, one must determine the spectral evolution of galaxy starlight 
photons from the IR through the UV range out to high redshifts. Pei \& Fall
(1995) have devised a clever method for calculating stellar emissivity as a 
function of redshift, one which is consistent with all recent data. We adopt 
this method and extend it by calculating the additional effect of metallicity 
evolution on stellar emissivity. We then calculate the $\gamma$-ray opacity 
of the universe to stellar photons at various redshifts and apply our results
to selected blazar spectra and the blazar background.

\section{CALCULATION OF STELLAR EMISSIVITY}
The basic idea of the Pei \& Fall (1995) approach, which we follow, is to 
relate the star formation rate to the evolution of neutral gas density in
damped Ly$\alpha$ systems and then to use the population synthesis models
(Bruzal \& Charlot 1993) 
to calculate the mean volume emissivity of the universe from stars as a 
function of redshift and frequency. Damped Ly$\alpha$ systems are believed
to be either the precursors to galaxies or young galaxies themselves. It is in 
these systems that initial star formation probably took place, so there is
a relationship between the mass content of stars and gas in these clouds. The
results obtained by Fall, \etal\ 1996 show excellent agreement with 
observational data obtained by the Canada-France redshift survey group for
redshifts out to 1 (Lilly, \etal\ 1996) and are consistent with lower limits
obtained on the emissivity at higher redshifts (Madau 1996). The stellar 
emissivity is found to peak between a redshift of 1 and 2 which is consistent 
with the results of ongoing observations from both the Hubble and Keck 
telescopes. 
We have made one significant
modification to the calculations of Fall, \etal\ (1996). We have attempted to
account for the significantly lower metallicity of early generation stars at 
higher redshifts which results in increased emission at shorter wavelengths 
and lowered emission at longer wavelengths. In order to estimate this effect,
we have used the results of Worthey (1994) and moderately extrapolated them
to both lower and higher wavelengths (Salamon \& Stecker 1997). We have also
considered the effect of dust opacity and have assumed a reasonable escape
factor to account for the fact that a small fraction of Lyman continuum photons
escape from galaxies unattenuated by stars and dust. The effect of this escape
factor on our subsequent opacity calculations is negligible, since there are 
not enough ionizing photons in intergalactic space to provide a significant
opacity to multi-GeV \grays.  Because the metallicity corrections
are less certain for the more massive stars ($M > 2M_{\odot}$), our 
metallicity-corrected UV radiation density should be viewed as an upper limit.

\section{OPACITY OF THE UNIVERSE TO GAMMA-RAYS}
Once the spectral energy density distribution of stellar photons in 
intergalactic space as a function of redshift is determined, the opacity
of the universe to \grays\ as a function of \gray\ energy and redshift can
be calculated (Stecker, \etal\ 1992). The basic processes which causes the
attenuation of \grays\ is the interaction of a \gray\ with a starlight photon
which results in the production of an electron-positron pair. Our results
indicate that \grays\ above an energy of $\sim$15 GeV will be attenuated if
they are emitted at redshifts greater than or equal to $\sim$3. The \gray\
burst observed by EGRET on 17 Feb 1994 contained a photon of energy $\sim$18 
GeV. Figure 1 shows the calculated opacity as a function of \gray\ 
energy for various source redshifts, with and without the metallicity
correction included; the true opacities
likely lie between the values shown in the left and right halves of Fig. 1.

\vskip 1.0truecm
\centerline{\psfig{figure=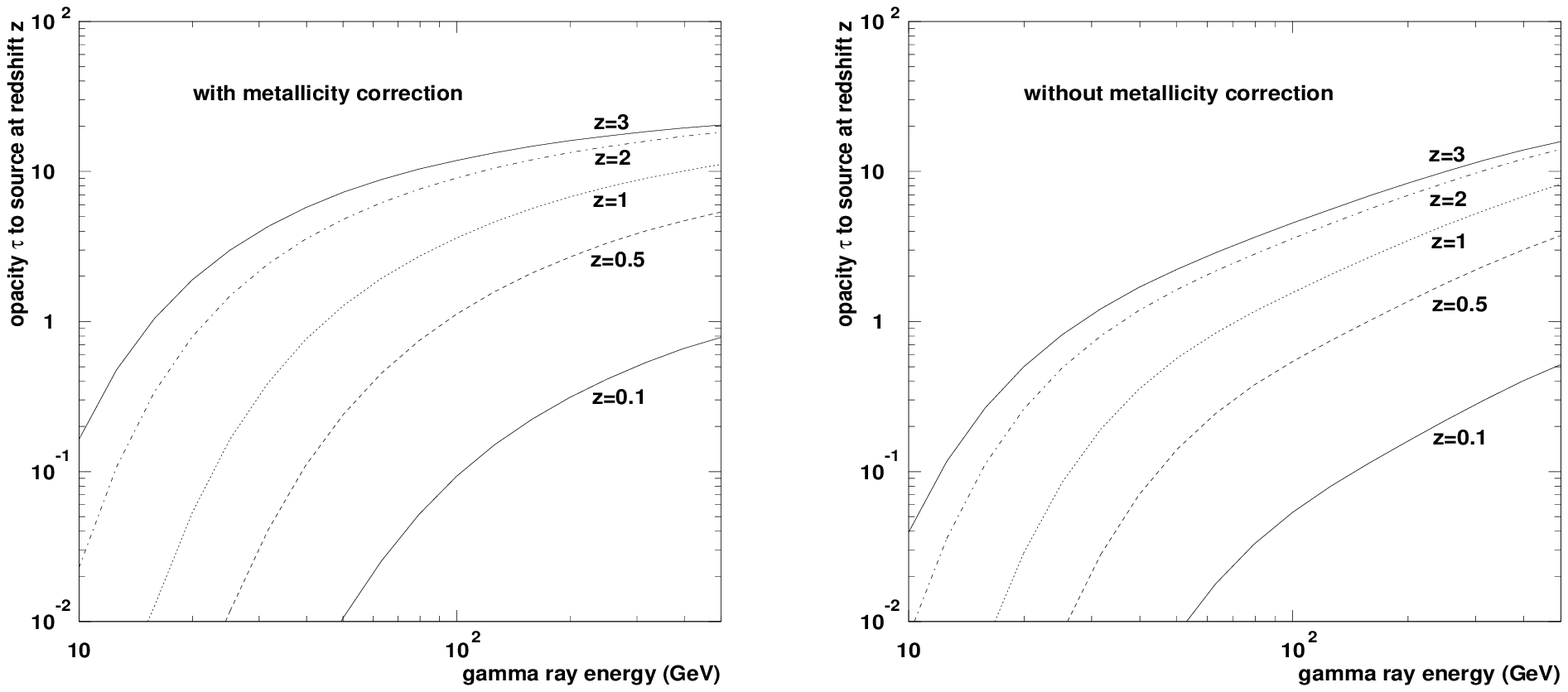,height=8.0truecm}\hskip 0.0truecm}
\vskip-0.0truecm
{\smallskip \it Figure 1: High energy \gray\ opacities calculated with and
without metallicity correction factor included.
\smallskip\smallskip}

Because the stellar emissivity peaks between a redshift of 1 and 2, there is
little increase in the \gray\ opacity when one goes to redshifts greater
than 2. This weak dependence indicates that the opacity is not determined
by the initial epoch of galaxy formation (which may be at $z \ge 5$), contrary
to the speculation of MacMinn \& Primack (1996).

\section{EFFECT OF ABSORPTION ON BLAZARS AND THE BACKGROUND}
Figure 2 shows the attenuation of \gray\ spectra resulting from
the opacities given in Figure 1 for the blazar sources 1633+382 
($z=1.81$), 3C279 ($z=0.54$), 3C273 ($z=0.15$), and Mrk421 ($z=0.031$).
The solid (dashed) lines result from
the opacities shown in left (right) half of 
Figure 1.  The redshift dependence of the break
energies is evident, as is the absence of any breaks below 10 GeV.
Future measurements of blazar spectral break energies will discriminate
between models of extragalactic extinction (such as this one) and those
involving cutoffs {\it intrinsic} to the source.

\vskip 1.0truecm
\centerline{\psfig{figure=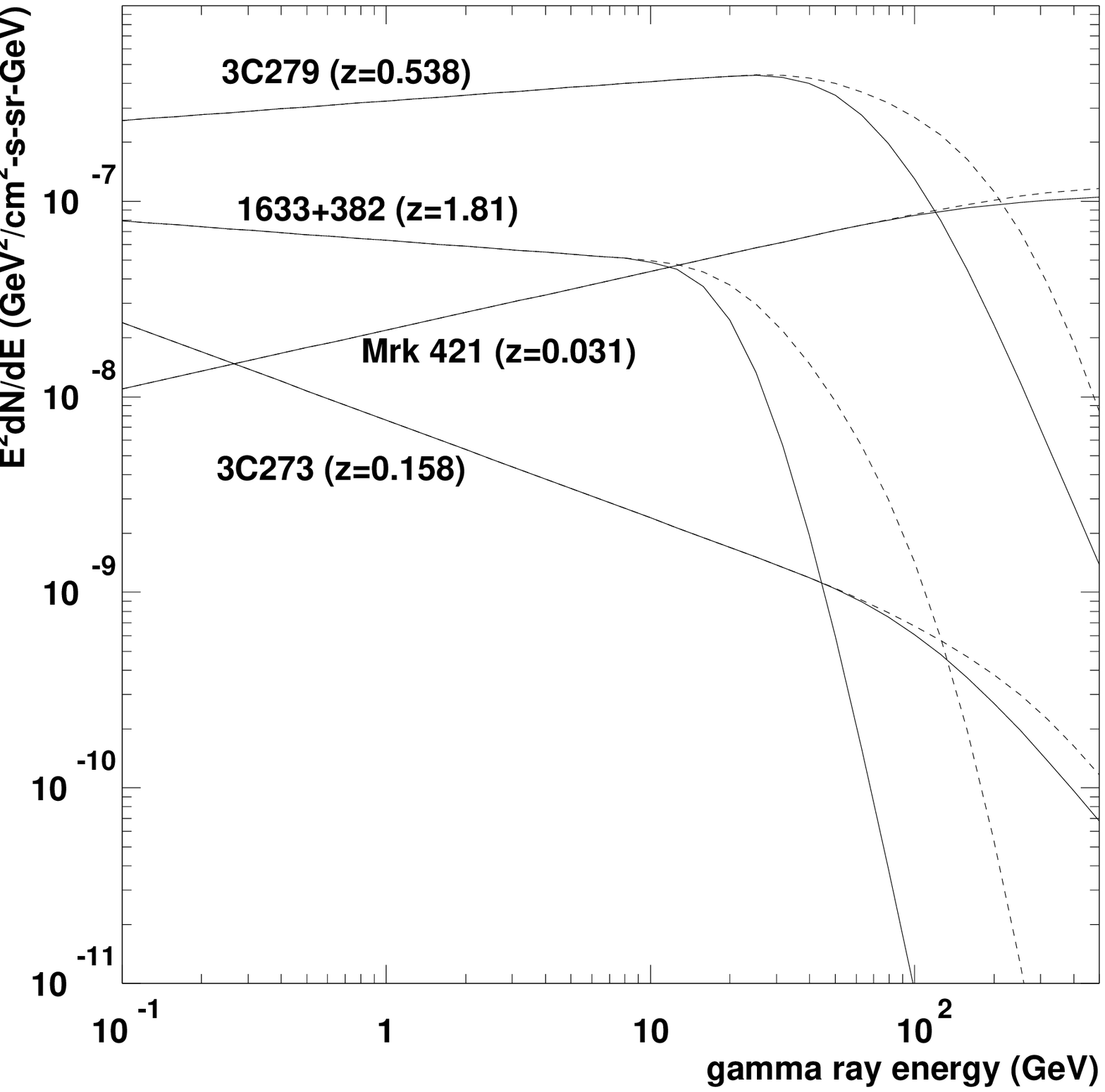,height=9.0truecm}\hskip 0.0truecm}
\vskip-0.0truecm
{\smallskip \it Figure 2: High energy \gray\ spectra of selected 
blazars attenuated
according to the opacities of Figure 1.  The solid (dashed) lines correspond
to the opacities calculated with (without) the metallicity correction.
\smallskip\smallskip}

Figure 3 shows the effect of absorption on the extragalactic \gray\ 
background computed using the unresolved blazar model of Stecker \& Salamon 
(1996). The solid (dashed) lines correspond to the metallicity correction being
included (neglected) in the opacity calculation.  The two families of curves
correspond to point source sensitivities of EGRET (top curves) and GLAST
(bottom curves).  (A better point source sensitivity results in the reduction 
in the number of {\it unresolved} \gray\ sources which contribute to
the \gray\ background.)  Also shown are the preliminary EGRET data on
the extragalactic \gray\ background spectrum (Fichtel 1996).  The cutoff
observed beyond $\sim$20 GeV reduces the effect of the extragalactic \gray\
background on searches for
\gray\ lines from neutralino-neutralino annihilation in the galactic
halo, should the extragalactic \gray\ background be due to unresolved 
flat-spectrum radio quasars.

\vskip 1.0truecm
\centerline{\psfig{figure=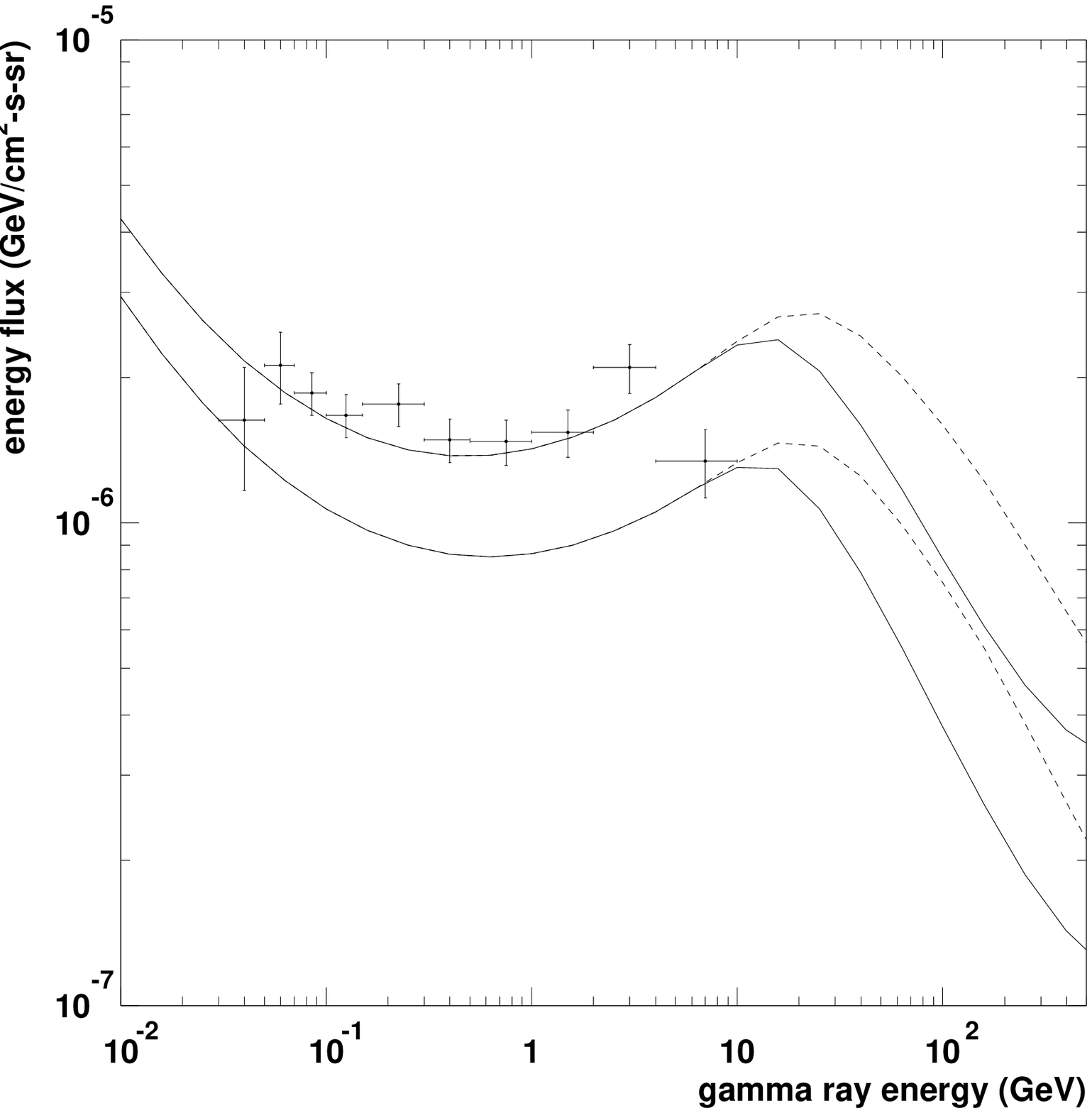,height=9.0truecm}\hskip 0.0truecm}
\vskip-0.0truecm
{\smallskip \it Figure 3: The extragalactic \gray\ background, calculated
with the model of Stecker and Salamon (1996), and attenuated with the
opacities of Fig. 1.  (See text.)
\smallskip\smallskip}

 \section{ACKNOWLEDGEMENTS}
We thank M. Fall, M. Malkan, Y. Pei, and G. Worthey for helpful
discussions and comments. 

\section{REFERENCES}
\setlength{\parindent}{-5mm}
\begin{list}{}{\topsep 0pt \partopsep 0pt \itemsep 0pt \leftmargin 5mm
\parsep 0pt \itemindent -5mm}
\vspace{-15pt}

\item Bruzal, A. G. \& Charlot, S. 1993, \apj\vol{405}{538}

\item Fall, S. M., Charlot, S. \& Pei, Y. C. 1996, \apj\vol{402}{479}

\item Fichtel, C. E. 1996, \aaps\vol{120}{23}

\item Lilly, \etal\ 1996, \apjl\vol{460}{L1}

\item MacMinn, D, \& Primack, J. 1966, \ssr\vol{75}{413}

\item Madau, P. \etal\ 1996 \mnras\vol{283}{138}

\item Pei, Y. C. \& Fall, S. M., 1995, \apj\vol{454}{69}

\item Salamon, M. H. \& Stecker, F. W. 1997, submitted to \apj 

\item Stecker, F.~W. \& De Jager, O.~C. 1997, \apj\vol{476}{712}

\item Stecker, F. W., De Jager, O. C. \& Salamon, M. H. 1992, 
\apjl\vol{390}{L49}

\item Stecker, F.W., and Salamon, M.H. 1996, \apj\vol{464}{600}

\end{list}

\end{document}